\documentclass{article}
\usepackage{graphicx}
\usepackage[english]{babel}
\usepackage[utf8]{inputenc}
\usepackage{graphicx}
\usepackage{amsmath}
\usepackage{amsfonts}
\usepackage{amssymb}
\usepackage[colorlinks]{hyperref}

\newtheorem{theorem}{Theorem}

\title{Fast {\large\it Cl-}type inhibitory neuron with delayed feedback has non-Markov output statistics}
\author{Alexander K.Vidybida\\
Bogolyubov Institute for Theoretical Physics\\
Metrologichna str., 14-B, Kyiv 03680, Ukraine
}

\begin{document}

\maketitle

\begin{abstract}
For a class of fast {\it Cl-}type inhibitory spiking neuron models with delayed feedback
 stimulated with a Poisson stochastic process of excitatory impulses, it is proven that
the stream of output interspike intervals cannot be presented as a Markov process of any order.
\end{abstract}

\section[Introduction]{Introduction
}

Spiking statistics of various neuronal models under a random stimulation has been studied
in the framework of two main approaches. The first one is named in \cite{Stein1967}
as ``Gaussian'', because it describes random stimulation by means of Gaussian noise,
see e.g. \cite{Bryant1976}.
This approach has developed into the well-known diffusion approximation methodology, see 
\cite{Capocelli1971}.
The second approach is named in  \cite{Stein1967} as ``quantal'', because it takes into account 
the discrete nature of the influence any input impulse may have on its target neuron.
The wide area of research    
 and applications    
 known as spiking neural networks, see 
\cite{Ghosh-Dastidar2009} for a review, could be considered as utilizing the quantal approach.

For a recent review of mathematically rigorous results regarding neuronal spiking statistics 
in the both approaches  see \cite{Sacerdote2013}.

 We study here mathematically
rigorously,  in the \linebreak framework of quantal approach, spiking
statistics of inhibitory neuron model belonging to a class of models 
(see Sec. \ref{Deass}, below)
with fast $Cl$-type inhibitory delayed feedback (see Fig. \ref{BNDwF}, below). 


\subsection[Biological inspiration]{Biological inspiration}\label{Bj}

Neurons, which send inhibitory impulses onto their own body or dendrites are known in real
nervous systems, see \cite{Bekkers1998,Bacci2003,Bacci2004,Smith2002}.

The chief inhibitory neurotransmitter in the \linebreak nervous system is 
Gamma-amino\-bu\-tyric acid (GABA). The GABA can activate several types of receptors, the main of which are
 GABA$_a$ and GABA$_b$. If GABA$_a$ receptors are activated, the excitable membrane
 becomes permeable for $Cl^-$ ions. If a neuron is partially excited, that is its
 membrane is depolarized, the $Cl^-$ current
 cancels this depolarization since the $Cl^-$ reversal potential is close/equal to the resting
 potential. For the same reason, the $Cl^-$ current 
 through open GABA$_a$ channels does not appear, if the membrane
 is at its resting potential.

 Another case is with GABA$_b$ receptors activation. This causes $K^+$ ions permeability.
The outward $K^+$ current is able to hyperpolarize membrane even below its resting potential.

The remarkable difference between GABA$_a$ and GABA$_b$ mediated inhibition is rather
different kinetics of the corresponding $Cl^-$ and $K^+$ currents. Namely, according to
     \cite{Benardo1994},
the $Cl^-$ current rise time is 1 - 5 ms, and the decay time constant is about 10 - 25 ms.
The $K^+$ current rise time is  10 - 120 ms, and the decay time constant is about 
200 - 1600 ms. The $K^+$ current can be even slower, see  \cite{Bacci2004,Storm1988a,Storm1990}.

Inspired by this contrast in the speed of $Cl^-$ and $K^+$ transients, we idealize
the $Cl^-$ current kinetics as having infinitesimally short rise time and infinitely
fast decay, both can be achieved with infinitely large $Cl^-$ conductance at the moment of receiving
inhibitory impulse. This kind of the $Cl^-$ current kinetics does ensure the perfect
reset of the membrane voltage to the resting state at the moment when inhibitory impulse arrives. 
Within the limited experimental data set
available for inhibitory autapses, see \cite{Benardo1994}, a single
impulse delivered through a single synapse ensures only partial reset. In this point,
our statement of the problem diverges from the current data. At the same time, 
in the artificial neuromorphic systems, see \cite{Sarpeshkar2009,Brderle2011}, the complete reset may well be realized.
Considering a partial reset would inappropriately increase the paper's dimensions.

\medskip

Below, it is expected that a neuron sends back to itself its output impulses through
a delayed feedback line, which ends with a GABA$_a$ autapse, see Prop1-Prop3, 
in Sec. \ref{line}.
This construction is stimulated with a Poisson stream of excitatory input impulses.
For this configuration it has been proven
in the previous paper \cite{Kravchuk2013} for the case of Poisson input stream and
for a concrete neuronal model --- the inhibitory binding neuron%
\footnote{Detailed description of the binding neuron model can be found in \cite{Vidybida2014}.\\ See also {\tt https://en.wikipedia.org/wiki/Binding\_neuron}.}
 with threshold 2 ---, 
 that statistics of its interspike intervals (ISIs) is essentially non-Markov\footnote{Sometimes, a concept of a point stochastic Markov process is
   confused with a process whose consecutive realizations are uncorrelated.
   Actually,
   the latter one is a renewal process, which is a specific case of
   Markov process. As regards a general Markov process, its
   realizations can well be correlated, see e.g. \cite{Lindner2010}. In this paper, we do not study
   correlations (which itself is an interesting topic, which could be
   addressed separately), but prove that the output statistics does not have
   the Markov property defined, e.g.,  in \cite[Ch.2 §6]{Doob1953}. Some
   interesting remarks about non-markoviannes can be found in
   \cite{vanKampen1998}.}. 
 In paper  \cite{Vidybida2015}, it has been proven for the Poisson input and 
 for a class of excitatory neuronal models that
 the presence of delayed feedback makes their activity non-Markov.
  In this paper, we use the approach developed in \cite{Vidybida2015} in order to refine 
and extend methods
of \cite{Kravchuk2013} making them applicable to any 
inhibitory neuron with fast {\it Cl}-type inhibition satisfying a number of  simple and
natural conditions, see Cond0-Cond4, below. 
The stimulation is assumed to be a Poisson stochastic
process.
Under those conditions, we prove rigorously that
ISI statistics of a neuron with delayed fast $Cl$-type inhibitory feedback
stimulated with a Poisson point stochastic process of input impulses
 cannot be represented as a Markov chain of any finite order. 
Finally, it should be mentioned that our consideration is valid also for artificial hardware
neurons, see  \cite{Rossell2012,Wang2013},
 and abstract neurons used in mathematical studies, provided that Cond0-Cond4 and Prop1-Prop3,
below, are satisfied.

\section{Definitions and assumptions}\label{Deass}

\subsection[Neuron without feedback]{Neuron without feedback}\label{neuron}

We assume that a neuron 
satisfies the following conditions:
\begin{itemize}
\item Cond0: Neuron is deterministic: Identical stimuli elicit identical spike trains from the same neuron.
\item Cond1: Neuron is stimulated with input Poisson stationary stream of excitatory impulses.
\item Cond2: Neuron may fire a spike only at a moment when it receives an input impulse.
\item Cond3: Just after firing, neuron appears in its resting state.
\item Cond4: The output interspike interval (ISI) distribution can be characterized with a
probability density function (pdf) $p^0(t)$, which is {\it continuous} with
\begin{equation}\label{nolj}
p^0(0)=0,
\end{equation}
positive: 
\begin{equation}\label{pos}
t>0\Rightarrow p^0(t)>0,
\end{equation}
 and bounded:
\begin{equation}\label{bound} 
\sup\limits_{t>0} p^0(t)< \infty.
\end{equation} 
By $t$ we denote the ISI's length.
Also, we impose on the function $p^0(t)$ the following condition: $t<0\Rightarrow p^0(t)=0$
in order to have it defined for all real numbers.
\end{itemize}

These conditions are, with some modifications, similar to those assumed in \cite{Vidybida2015} 
for a class of excitatory neurons.
The modifications are as follows:

Cond3 ---  we assume that after firing a neuron appears in its
resting state with all excitation canceled, while in \cite{Vidybida2015} it is a standard state, which
not necessarily is the resting one. 

Cond4 --- the requirement of continuity
of $p^0(t)$
is added as compared to \cite{Vidybida2015}. This addition has a pure mathematical nature and
seems to be valid for any ``good'' neuronal model (without feedback). The subsequent proof
of non-markoviannes relies on it.

The Cond3, above, limits the set of models as compared to \cite{Vidybida2015}. Namely, it claims
that the standard state of \cite[Cond3]{Vidybida2015} 
must be exactly the resting state of neuron. This requirement
is imposed due to the specifics of {\em Cl-}type fast inhibition. 
 For our approach, it is important that after receiving inhibitory impulse,
the neuron appears in exactly 
the same state as immediately after firing. And the state after receiving 
{\em Cl-}type  inhibitory impulse can be only the resting state, see Sec. \ref{Bj}, above.

It seems that these conditions are satisfied for many threshold-type neuronal models known in the
literature, see  \cite{Burkitt,Chacron2003,Jolivet2004,Jolivet2006} and citations therein.
But this still has to be proven by calculating corresponding $p^0(t)$.
At least, all the five conditions are satisfied for the binding neuron model and for the basic 
leaky integrate-and-fire
(LIF) model, both for Poisson stimulation. See  \cite{Vidybida2007,Vidybida2014b}, where $p^0(t)$ is calculated exactly for each model, respectively.

\subsection{Feedback line action}\label{line}

We expect that the feedback line satisfies Prop1, Prop2 of \cite{Vidybida2015}, 
which are reproduced below for completeness.
The Prop3  of \cite{Vidybida2015} should be modified for the {\em Cl-}type fast inhibition
as shown below:
\begin{itemize}
\item Prop1: The time delay in the line is $\Delta>0$.
\item Prop2: The line is able to convey no more than one impulse.
\item Prop3: The impulse conveyed to the neuronal input is the fast $Cl$-type inhibitory impulse.
This means that after receiving such an impulse, the neuron appears in its resting state.
 This exhausts the action of the inhibitory impulse
in a sense that it has no influence on further neuronal states created by next 
excitatory impulses. It as well does not affect neuron 
being in its resting state.
\end{itemize}

The Prop1 expects that the delay is always the same and each impulse,
entered the line is delivered to its output and effects the neuron.
Thus, we do not consider here cases when the transmission is unreliable, or
the delay time is not constant.

The validity of the Prop2 depends on relation between the conduction velocity, recovery time
and the line's length. Also Prop2 seems plausible if the firing frequency is low.

The Prop3 just characterizes a neuronal model as inhibitory with GABA$_a$-type autapse.
Its validity depends on the fact that the $Cl^-$ reversal potential is identical to
the resting potential. In same cases this is fulfilled, see \cite{Benardo1994}.
It is also expected that a single AP delivered by a feedback line is potent enough
for canceling any excitation present. Taking into account that observed single GABA$_a$ IPSP
peak value rare exceeds 6 mV, this may require the delay line sprouting into several
autaptic endings.

\begin{figure} [h]
\begin{center}
        \includegraphics[width=0.69\textwidth]{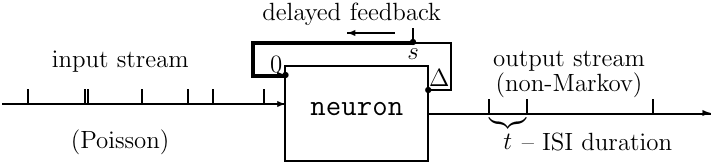}
\end{center}
\caption{\label{BNDwF} Neuron with delayed feedback. As {\large\tt neuron} in the figure,
we consider any neuronal model, which satisfies the set of conditions Cond0 - Cond4, above.}
\end{figure}

The important for us consequence of the Prop2, above, is that at any moment of time the feedback
line is either empty, or conveys a single impulse. If it does convey an impulse, then its state
can be described with a stochastic variable $s$, $s\in ]0;\Delta]$, which we call further ``time to live'',
see Fig. \ref{BNDwF}. 
The variable $s$ denotes the exact time required by the impulse to reach the output end of the line,
which is the neuron's input
for inhibitory impulses, and to leave the line with the consequences
described in the Prop3, above. Here, $\Delta$ denotes the delay duration in the
feedback line. Notice, that at the beginning of any ISI, the line is never empty.

\section{Results}

Our purpose here is to prove the following Theorem\footnote{A similar theorem for the {\em excitatory}
feedback line has been proven in \cite{Vidybida2015}.}:
\begin{theorem}
Let a neuronal model satisfies conditions Cond0-Cond4, above.
Suppose that the model is extended by introducing a delayed { fast {\it Cl}-type
 inhibitory} feedback line,
which satisfies the Prop1-Prop3, above. Then, in the stationary regime,
the output stream of ISIs of the neuron cannot be presented as a Markov chain
of any finite order.
\end{theorem}

\subsection[Proof outline]{Proof outline
}

Let $p^{inh}(t_{n+1}\mid t_n,\dots,t_0)dt_{n+1}$ denote the conditional probability 
to get the duration of $(n+2)$-nd ISI in the interval
$[t_{n+1};t_{n+1}+dt_{n+1}[$ provided 
that previous $n+1$ ISIs have duration $t_n,\dots,t_0$, respectively.
From the definition in  \cite[Ch.2 §6]{Doob1953},
one can obtain the necessary condition
\begin{equation}
\label{def}
        p^{inh}(t_{n+1}\mid t_{n},\ldots,t_{1},t_{0})
        = p^{inh}(t_{n+1}\mid t_{n},\ldots,t_{1}),~~
\end{equation}
required for the stochastic process $\{t_{j}\}$ to be $n$th order Markov chain.
Notice, that (\ref{def}) must be satisfied for any values of the variables
$t_i$, $i=0,\ldots,n+1$.

We intend to prove that the relation (\ref{def}) does not hold for any $n$. For this purpose
we calculate exact expression for $p^{inh}(t_{n+1}\mid t_n,\dots,t_0)$ as
\begin{equation}\label{defcond}
p^{inh}(t_{n+1}\mid t_n,\dots,t_0)=
\frac{p^{inh}(t_{n+1}, t_n,\dots,t_0)}{p^{inh}(t_n,\dots,t_0)}
\end{equation}
from which it will be clearly seen that the $t_0$-dependence in $p^{inh}(t_{n+1}\mid t_n,\dots,t_0)$
cannot be eliminated whatever large the $n$ is. 
In the Eq. (\ref{defcond}), expression $p^{inh}(t_n,\dots,t_1)$ denotes the 
joint probability density function of ISIs duration of neuron with the fast {\em Cl-}type inhibitory delayed feedback.

Let us introduce the conditional joint probability density
$p^{inh}(t_{n+1},\dots,t_0\mid s)$, which denotes the conditional probability density to get
$n+2$ consecutive ISIs $\{t_{n+1},\dots,t_0\}$ provided that at the beginning of the first ISI ($t_0$)
the time to live of impulse in the feedback line is equal to $s$.
This conditional probability can be
used to calculate required joint pdfs as follows
\begin{equation}
\label{pdfs}
p^{inh}(t_{n+1},\dots,t_0)=
\int\limits_0^\Delta p^{inh}(t_{n+1},\dots,t_0\mid s)f^{inh}(s)\,ds,
\end{equation}
where 
$f^{inh}(s)$ is the stationary pdf which describes distribution of times to live
at the beginning of any ISI in the stationary regime.

In what follows we analyze the structure of functions
$f^{inh}(s)$ and $p^{inh}(t_{n+1},$ $\dots,t_0\mid s)$. It appears that $f^{inh}(s)$ has a singular
component $a\delta(s-\Delta)$ with $a>0$, and $p^{inh}(t_{n+1},\dots,t_0\mid s)$
has jump discontinuities at definite hyper-planes in the $(n+3)$-dimensional
space of its variables $(t_{n+1},\dots,t_0,s)$. 
 After integration in (\ref{pdfs}), some of those discontinuities will
survive in the $(n+2)$-dimensional space of variables $(t_{n+1},\dots,t_0)$,
and exactly one of those survived has its position depending on $t_0$. 
The $t_0$-dependent jump discontinuity
will as well survive in the $p^{inh}(t_{n+1}\mid t_n,\dots,t_0)$ for any $n$,
provided that $t_{n},\dots,t_0$ satisfy the following condition:
\begin{equation}\label{Domain}
\sum\limits_{i=0}^n t_i<\Delta,
\end{equation} 
where $\Delta>0$ is the full delay time in the feedback line.
Taking into account that the equation in the necessary condition (\ref{def}) must hold
for any set of $t_{n+1},\dots,t_0$, we conclude that (\ref{def}) cannot be satisfied for
any $n$.

\subsection{The proof}

\subsubsection{Structure of functions $p^{inh}(t_{n+1},\dots,t_0 \mid s)$}

Specifics of the feedback line action together with condition (\ref{Domain}) results
in a very simple structure of \linebreak $p^{inh}(t_{n+1},\dots,t_0\mid s)$ at different parts of the 
integration domain in (\ref{pdfs}). Those parts are defined as follows:

\begin{equation*}
D_k=\{s\mid \sum\limits_{i=0}^{k-1}t_i<s \le \sum\limits_{i=0}^{k}t_i\},\,
k=0,\dots,n,
\quad
D_{n+1}=\{s\mid \sum\limits_{i=0}^{n}t_i <s \le \Delta\}\,.
\end{equation*}
As regards the structure itself, the following representation can be derived
similarly as it was done in \cite{Vidybida2015}:
\begin{multline}\label{struk}
p^{inh}(t_{n+1},\dots,t_0 \mid s)=
p^{inh}(t_{n+1},\dots,t_{k+1} \mid \Delta)\times\\\times
p^{inh}\left(t_k \mid s-\sum\limits_{i=0}^{k-1} t_i\right) 
\prod\limits_{i=0}^{k-1}p^0(t_i),
\\
s\in D_k,\quad k=0,\dots,n,
\end{multline}
\begin{multline}
\label{strun+1}
p^{inh}(t_{n+1},\dots,t_0 \mid s)=
\\=
p^{inh}\left(t_{n+1} \mid s-\sum\limits_{i=0}^{n} t_i\right)
\prod\limits_{i=0}^{n}p^0(t_i),\quad
 s\in D_{n+1}.
\end{multline}
\begin{multline}
\label{strukf}
p^{inh}(t_{n+1},\dots,t_{k+1} \mid \Delta)=
\\
=p^{inh}\left(t_{n+1} \mid \Delta-\sum\limits_{i=k+1}^{n} t_i\right)
\prod\limits_{i=k+1}^{n}p^0(t_i).
\end{multline}
Here $p^{inh}(t\mid s)$ denotes the conditional pdf to get ISI of duration $t$ if
at its beginning, time to live of impulse in the feedback line is $s$. 

Representation of $p^{inh}(t_{n+1},\dots,t_0 \mid s)$ by means of $p^0(t)$ and $p^{inh}(t\mid s)$ 
found here is
similar to that found in  \cite{Vidybida2015} for the excitatory case. 
But the structure of function $p^{inh}(t \mid s)$, used in that representation, is different.

\subsubsection{Structure of function $p^{inh}(t \mid s)$}\label{Spts}

Expect that at the beginning of an ISI, there is an impulse in the 
feedback line with time to live $s$. Then the probability that this ISI
will have its duration $t<s$ does not depend on the feedback
line presence. Therefore,
\begin{equation}\label{pinit}
t<s\, \Rightarrow\, p^{inh}(t\mid s) = p^0(t).
\end{equation}

In the opposite situation, receiving of an ISI duration greater than $s$
happens if (i) the neuron is not firing during interval $]0;s[$
and (ii) the neuron starts at its resting state (Prop3, above) at the moment $s$
and fires at $t>s$. Realizations of events (i) and (ii) depend on
 disjoint segments of the input Poisson stream (Cond1, above).
Therefore, (i) and (ii) are statistically independent. The probability of (i)
is as follows:
\begin{equation}\label{P0}
\mathbf{P}^0(s) = 1 - \int\limits_0^{s} p^0(t)dt.
\end{equation}
The probability of (ii) is $p^0(t-s)$. This gives
\begin{equation}\label{pisnit}
t>s\, \Rightarrow\, p^{inh}(t\mid s) = \mathbf{P}^0(s)p^0(t-s).
\end{equation}
It can be concluded from (\ref{pinit}) and (\ref{pisnit}) that
$$
\lim\limits_{t\uparrow s} p^{inh}(t\mid s) = p^0(s)\quad
\text{and}\quad
\lim\limits_{t\downarrow s} p^{inh}(t\mid s) = 0.
$$
Now, taking into account (\ref{nolj}) and (\ref{pos}) from Cond4, above, we conclude that the function 
$p^{inh}(t\mid s)$
considered as a function of two variables $(t,s)$, $t\ge0$, $s\in\,]0;\Delta]$ has a jump
discontinuity along the straight line $t=s$. The magnitude of this jump is 
$p^0(s)$, and it is strictly
positive for positive $t$. Concrete values of $p^{inh}(t\mid s)$ along the line $t=s$ does not matter and can be chosen 
arbitrarily.

Finally, for $p^{inh}(t\mid s)$ we have%
\footnote{Compare this with  \cite[Eq. (11)]{Vidybida2013}, where $p^{inh}(t\mid s)$ is 
calculated exactly for the binding neuron model stimulated with Poisson stream.}
\begin{equation}\label{pts}
p^{inh}(t\mid s) = \chi(s-t) p^0(t) + \mathbf{P}^0(s)p^0(t-s),
\end{equation}
where $\chi(s)$ is the Heaviside step function.

\subsubsection{Structure of probability density function $f^{inh}(s)$}

Everywhere in this paper we expect that all pdfs $p^{inh}(t_{n+1},\dots,t_0)$
have achieved their stationary form, and we analyze the stationary regime.
But any stationary regime arises from some initial distribution.
In principle, different initial distributions may result in different 
final stationary distributions.

As it can be concluded from (\ref{struk})-(\ref{strukf}), the only quantity,
which might depend on initial conditions in the right-hand side of representation
(\ref{pdfs}) is the pdf $f^{inh}(s)$.

Before the stationary regime is achieved, $f^{inh}(s)$ is changed after each firing:
\begin{equation}\label{transtep}
f_{n+1}(s) = \int\limits_0^\Delta \mathbf{P}(s\mid s') f_n(s') ds',
\end{equation}
where the transition function $\mathbf{P}(s\mid s')$ gives the probability density
to find at the beginning of an ISI an impulse in the line with time to live $s$
provided that at the beginning of the previous ISI, there was an impulse with time to live $s'$.
In the stationary regime, the pdf $f(s)$ must satisfy the following equation
\begin{equation}\label{trans}
f^{inh}(s) = \int\limits_0^\Delta \mathbf{P}(s\mid s') f^{inh}(s') ds',
\end{equation}
Now, the question of existence and uniqueness of the stationary regime
might be resolved by analyzing Eqs. (\ref{transtep}) and (\ref{trans})
for convergence and uniqueness. 
This is expected to do in another paper.
In this paper we assume that sequence $\{f_n^{inh}(s)\}$
of pdfs generated by Eq. (\ref{transtep}) converges to some pdf for any initial $f_0(s)$, 
and admit that there might be different limiting distributions for different
$f_0(s)$. 

The exact expression for $\mathbf{P}(s\mid s')$ is found in \cite[Eqs.(11)-(13)]{Vidybida2015}.
It appears that the structure of $f^{inh}(s)$, which follows from (\ref{trans}) is exactly the same as it has been found 
in \cite{Vidybida2015} for the excitatory case. This structure is as follows%
\footnote{Compare this with  \cite[Eqs. (14)-(16)]{Vidybida2008a}, where $f(s)$ is calculated exactly for the binding neuron model.}
\begin{equation}\label{ff}
f^{inh}(s)=g(s) + a\delta(s-\Delta),
\end{equation}
where $a>0$ and $g(s)$ is bounded continuous function vanishing out of interval $]0;\Delta[$.

\subsubsection{Form of $p^{inh}(t_{n+1},\dots,t_0)$ and $p^{inh}(t_{n},\dots,t_0)$ after integration in (\ref{pdfs})} 

Let $D=\bigcup\limits_{k=0}^n D_k$. 
At $D$, representations (\ref{struk}) and (\ref{strukf}) are valid. Also at $D$,
$f^{inh}(s)$ reduces to $g(s)$.
Therefore,
\begin{multline}\label{intD}
\int\limits_D p^{inh}(t_{n+1},\dots,t_0\mid s) f^{inh}(s)\, ds =
\\=
\sum\limits_{k=0}^np^{inh}\left(t_{n+1}\mid \Delta -\sum\limits_{i=k+1}^n t_i\right)\times
\\\times
\prod\limits_{\vbox{\footnotesize\hbox{$i=0$}\hbox{$i\ne k$}}}^np^0(t_i)
\int\limits_{D_k}p^{inh}\left(t_k\mid s-\sum\limits_{j=0}^{k-1} t_j\right) g(s)ds.
\end{multline}
The first factor (with fixed $k$,  $0\le k\le n$) in the r.h.s. of Eq. (\ref{intD}) is as follows:
$$
p^{inh}\left(t_{n+1}\mid \Delta -\sum\limits_{i=k+1}^n t_i\right).
$$
Due to Eq. (\ref{pts}), this factor does have a jump discontinuity along the hyperplane
$
\sum\limits_{i=k+1}^{n+1}t_i=\Delta
$
in the space of variables $(t_0,\dots,t_{n+1})$. Notice, that the position of this hyperplane
does not depend on $t_0$ for any $k\in\{0,\dots,n\}$.

The second factor in the r.h.s. of Eq. (\ref{intD}) is as follows:
$
\prod\limits_{\vbox{\footnotesize\hbox{$i=0$}\hbox{$i\ne k$}}}^np^0(t_i),
$
and it is continuous.

The third factor in the r.h.s. of Eq. (\ref{intD}) can be transformed as follows:
\begin{multline}\label{ThirdFactor}
\int\limits_{D_k}p^{inh}\left(t_k\mid s-\sum\limits_{j=0}^{k-1} t_j\right) g(s)ds=
\\=
\int\limits_{\sum\limits_{j=0}^{k-1} t_j}^{\sum\limits_{j=0}^{k} t_j}
p^{inh}\left(t_k\mid s-\sum\limits_{j=0}^{k-1} t_j\right) g(s)ds=
\\
=\int\limits_0^{t_k}p^{inh}(t_k\mid s) g\left(s+\sum\limits_{j=0}^{k-1} t_j\right)ds=
\\=
\int\limits_0^{t_k}
\mathbf{P}^0(s)
p^0(t_k- s)
 g\left(s+\sum\limits_{j=0}^{k-1} t_j\right)ds.
 \end{multline}
The last expression is continuous with respect to variables $(t_0,\dots,t_{n+1})$.
Therefore, one can conclude that expression (\ref{intD}) does not have a jump discontinuity,
which position depends on $t_0$.

Consider now the remaining part of integral in (\ref{pdfs}). With (\ref{strun+1})
taken into account one has:
\begin{multline}\label{intDn+1}
\int\limits_{D_{n+1}} p^{inh}(t_{n+1},\dots,t_0\mid s) f^{inh}(s)\, ds
=
\\=
\prod\limits_{i=0}^n p^0(t_i)
\int\limits_{D_{n+1}}p^{inh}\left(t_{n+1}\mid s-\sum\limits_{i=0}^{n}t_i\right) f^{inh}(s)ds.
\end{multline}
Here, the first factor, $\prod\limits_{i=0}^n p^0(t_i)$ is continuous and strictly positive
for positive $t_i$. The second factor can be transformed as follows:
\begin{multline}\label{2fak}
\int\limits_{D_{n+1}}p^{inh}\left(t_{n+1}\mid s-\sum\limits_{i=0}^{n}t_i\right) f^{inh}(s)ds
=
\\=
\int\limits_{\sum\limits_{i=0}^{n}t_i}^\Delta
p^{inh}\left(t_{n+1}\mid s-\sum\limits_{i=0}^{n}t_i\right) f^{inh}(s)ds=
\\=
\int\limits_0^{\Delta-\sum\limits_{i=0}^{n}t_i}
p^{inh}(t_{n+1}\mid s) f^{inh}\left(s+\sum\limits_{i=0}^{n}t_i\right)ds.
\end{multline}
Now, let us use representations (\ref{pts}) and (\ref{ff}) in order to figure out
which kind of discontinuities does the expression (\ref{2fak}) have. 
Due to (\ref{pts}) and (\ref{ff}), expression (\ref{2fak}) will have four terms.
The first one we get by choosing the first term both in (\ref{pts}) and (\ref{ff}):
$$
A_{11}=\int\limits_0^{\Delta-\sum\limits_{i=0}^{n}t_i}
\chi(s-t_{n+1}) p^0(t_{n+1})
g\left(s+\sum\limits_{i=0}^{n}t_i\right)ds.
$$
This term is either equal to zero, if $t_{n+1}>\Delta-\sum\limits_{i=0}^{n}t_i$,
or otherwise transforms into a continuous function of variables $(t_0,\dots,t_{n+1})$. Moreover, 
$$\lim\limits_{t_{n+1}\uparrow \Delta-\sum\limits_{i=0}^{n}t_i} A_{11}(t_{n+1})=0.$$

The second one we get by choosing the second term in (\ref{pts}) and the first term in (\ref{ff}):
$$
A_{21}=
\int\limits_0^{\Delta-\sum\limits_{i=0}^{n}t_i}
\mathbf{P}^0(s)
p^0(t_{n+1} - s)
g\left(s+\sum\limits_{i=0}^{n}t_i\right)ds.
$$
This is as well a continuous function of variables\linebreak $(t_0,\dots,t_{n+1})$.

The third one we get by choosing the first term in (\ref{pts}) and the second term in (\ref{ff}):
\begin{multline}\label{third}
A_{12}=
\\=
a\int\limits_0^{\Delta-\sum\limits_{i=0}^{n}t_i}
\chi(s-t_{n+1}) p^0(t_{n+1})
\delta\left(\sum\limits_{i=0}^{n}t_i +s  - \Delta\right)\,ds=
\\
=a \chi\left(\Delta-\sum\limits_{i=0}^{n+1}t_i\right)p^0(t_{n+1}).
\end{multline}
This term has a jump discontinuity along the hyperplane
\begin{equation}\label{hyper}
\sum\limits_{i=0}^{n+1}t_i = \Delta\,.
\end{equation}

The forth one we get by choosing the second term in (\ref{pts}) and the second term in (\ref{ff}):
\begin{multline}\nonumber
A_{22}=
\\=
a\int\limits_0^{\Delta-\sum\limits_{i=0}^{n}t_i}
\mathbf{P}^0(s)
p^0(t_{n+1} - s)
\delta\left(\sum\limits_{i=0}^{n}t_i +s -\Delta\right)\,ds=
\\=
\mathbf{P}^0\left(\Delta-\sum\limits_{i=0}^{n}t_i\right)
p^0\left(\sum\limits_{i=0}^{n+1}t_i - \Delta\right).
\end{multline}
This is as well a continuous function of variables\linebreak $(t_0,\dots,t_{n+1})$.

After taking into account the above reasoning, we conclude that the required joint
probability density has the following form
\begin{equation}
\label{pn+1}
p^{inh}(t_{n+1},\dots,t_0)=
p^w(t_{n+1},\dots,t_0)+
a\chi\left(\Delta-\sum\limits_{i=0}^{n+1}t_i\right)
\prod\limits_{j=0}^{n+1} p^0(t_j).
\end{equation}
where function $p^w(t_{n+1},\dots,t_0)$ does not have a jump discontinuity depending on $t_0$, and the second term in (\ref{pn+1}) does have such a 
discontinuity along the hyperplane (\ref{hyper}).\bigskip

\noindent
{\it Form of $p^{inh}(t_{n},\dots,t_0)$ after integration}\label{Form}\medskip

\noindent
If (\ref{Domain}) is satisfied, then we have similarly to (\ref{struk}), (\ref{strun+1})
\begin{multline}\nonumber
p^{inh}(t_{n},\dots,t_0 \mid s)=p^{inh}(t_{n},\dots,t_{k+1} \mid \Delta)\times\\\times
p^{inh}\left(t_k \mid s-\sum\limits_{i=0}^{k-1} t_i\right)
\prod\limits_{i=0}^{k-1}p^0(t_i),
\\
 s\in D_k,\quad k=0,\dots,n-1,
\end{multline}
\begin{equation}\nonumber
p^{inh}(t_{n},\dots,t_0 \mid s)=
p^{inh}\left(t_n \mid s-\sum\limits_{i=0}^{n-1} t_i\right)
\prod\limits_{i=0}^{n-1}p^0(t_i),\quad s\in D_n.
\end{equation}
Again due to (\ref{Domain}), and in analogy with (\ref{strukf}) we have instead 
of the last two equations the following one:
\begin{multline}
\label{strn}
p^{inh}(t_{n},\dots,t_0 \mid s)=
p^{inh}\left(t_k \mid s-\sum\limits_{i=0}^{k-1} t_i\right)
\prod\limits_{\vbox{\footnotesize\hbox{$i=0$}\hbox{$i\ne k$}}}^{n}p^0(t_i),\quad
\\
 s\in D_k,\,\, k=0,\dots,n.
 \end{multline}
It is clear that expression similar to (\ref{strun+1}) turns here into the following
\begin{equation}\label{strn+1}
p^{inh}(t_{n},\dots,t_0 \mid s)=
\prod\limits_{i=0}^{n}p^0(t_i),\quad  s\in D_{n+1}.
 \end{equation}
Now, due to (\ref{strn}), (\ref{strn+1}) we have
\begin{multline}\label{pn}
p^{inh}(t_{n},\dots,t_0)=\int\limits_0^\Delta p^{inh}(t_{n},\dots,t_0 \mid s)f^{inh}(s)ds=
\\=
\sum\limits_{k=0}^n
\prod\limits_{\vbox{\footnotesize\hbox{$i=0$}\hbox{$i\ne k$}}}^{n}p^0(t_i)
\int\limits_{D_k}
p^{inh}\left(t_k \mid s-\sum\limits_{i=0}^{k-1} t_i\right)
g(s) ds+
\\+
\prod\limits_{i=0}^{n}p^0(t_i)\int\limits_{D_{n+1}}f^{inh}(s) ds.
\end{multline}
From calculations similar to those made in Eq. (\ref{ThirdFactor}) it can be concluded
that $p^{inh}(t_{n},\dots,t_0)$ is continuous at the domain defined by (\ref{Domain}).

\subsubsection{$t_0$-dependence cannot be eliminated in $p^{inh}(t_{n+1} \mid t_n,\dots,t_0)$}

Now, with representations (\ref{pn+1}) for $p^{inh}(t_{n+1},\dots,t_0)$ and
(\ref{pn}) for $p^{inh}(t_{n},\dots,t_0)$ we can pose a question about the form
of $p^{inh}(t_{n+1}\mid t_n,\dots, t_0)$. The latter can be found as defined in (\ref{defcond}).
First of all notice that due to (\ref{pn}) and Cond4,
$p^{inh}(t_{n},\dots,t_0)$ is strictly positive for positive ISIs. 
This allows us to use it as denominator in the definition (\ref{defcond}).
Second, it can be further concluded from (\ref{pn}) and Cond4, that $p^{inh}(t_{n},\dots,t_0)$ is
bounded.
The latter together with continuity of $p^{inh}(t_{n},\dots,t_0)$
means that any discontinuity of jump type present in the $p^{inh}(t_{n+1},\dots,t_0)$
appears as well in the $p^{inh}(t_{n+1}\mid t_n,\dots, t_0)$. It follows from the above 
and from Eq. (\ref{pn+1})
that the conditional pdf $p^{inh}(t_{n+1}\mid t_n,\dots, t_0)$ can be represented in the
following form:
\begin{multline}\label{firepr}
p^{inh}(t_{n+1}\mid t_n,\dots, t_0) = 
p^w(t_{n+1}\mid t_n,\dots, t_0)+
\\+
Z(t_{n+1},\dots,t_0)\chi\left(\Delta-\sum\limits_{i=0}^{n+1}t_i\right),
\end{multline}
where $p^w(t_{n+1}\mid t_n,\dots, t_0)$ does not have any jump type discontinuity
which position
depends on $t_0$, and $Z(t_{n+1},\dots,t_0)$ is strictly positive function:
$$
Z(t_{n+1},\dots,t_0)=
\frac{a\prod\limits_{i=0}^{n+1}p^0(t_i)}{p(t_{n},\dots,t_0)}.
$$
Thus the representation (\ref{firepr}) proves  that for any $n$,
conditional pdf $p^{inh}(t_{n+1}\mid t_n,\dots, t_0)$ does depend on $t_0$ 
(the second term in (\ref{firepr})) 
and this dependence cannot be eliminated.
$\Box$

See also Appendix, below, where the above general reasoning is illustrated
for the LIF neuronal model with threshold 2 
(that is two input impulses applied in a short succession are able to trigger,
see (\ref{Th=2}), below).

\section{Discussion and Conclusions}

The question as to what extent the 
stream of neuronal output impulses can be modeled as Poisson stream
 has been discussed in neuroscience, see \cite{Averbeck2009}.
The experimentally observed presence of memory in the ISIs output of real neurons has been
reported many times, see  \cite{Lowen1992,Shinomoto1999,Ratnam2000,Nawrot2007,Maimon2009}.
Also several theoretical models of how the memory could appear are offered, 
see \cite{Chacron2003,Rospars1993a,Lnsk1999,Benedetto2013,Kass2005,Avila-Akerberg2011,Cessac2011}.

In this paper we use the quantal approach, as it is defined in \cite{Stein1967} in order to 
prove that the Markov property is broken in the ISI output stream of a neuronal model belonging
to a defined class of models, equipped with delayed fast {\em Cl}-type inhibitory feedback,
which is stimulated with a Poisson stochastic process of input excitatory impulses.
Several previous results obtained in the quantal approach are used in this paper,
see \cite{Kravchuk2013,Vidybida2015,Vidybida2007,Vidybida2013,Vidybida2008a}.

In all these papers (as in many other computational neuroscience works) it is assumed
that input impulse has zero duration and is modeled as the Dirac $\delta$-function in time.
On the other hand, the data of papers \cite{Benardo1994,Bacci2003}, which inspired this study,
suggest that observed there width of inhibitory impulse is comparable with or even longer than
 the delay time for 
autaptic selfinhibition. In the case of that extended impulses it is not clear which figure
should be considered as the delay. In this connection, it should be taken into account that
any extended in time impulse can be represented as a sum of short impulses precisely positioned
at different moments of time. For each short impulse component the delay value,
either axonal, or synaptic, or both is precisely
defined. Of course, we cannot expect here that each short impulse component performs
a complete reset of membrane potential. Therefore, additional analysis is required,
which is out of scope of this paper. On the other hand, for neurocybernetical artificial devices,
see, e.g.  \cite{Sarpeshkar2009,Brderle2011}, situation with short impulse and long delay
seems to be more natural due to specifics of digital devices.

The first results this paper is based on are obtained for the binding neuron (BN) model.
Namely, in \cite{Vidybida2007} under Poisson stimulation the output ISI pdf and mean ISI are obtained
for the BN with threshold (Th) 2, and the mean ISI for Th = 3. 
In \cite{Kravchuk2013,Vidybida2013,Vidybida2008a} a BN model with Th = 2 and with delayed feedback,
either excitatory or inhibitory, stimulated with Poisson stream is considered. For this case, the ISI pdf is 
found and also it is proven that the output ISI stream is non-Markov. In \cite{Vidybida2015},
any neuronal model from a defined class is considered. A delayed feedback is assumed {\em excitatory}, 
and stimulation is Poissonian. For this case, it is proven that the output stream is non-Markov.

In this paper, a class of neuronal models with delayed {\em Cl}-type {\em inhibitory} feedback is considered.

The memory property in output ISI streams is often discussed in terms of correlation coefficient (CC),
e.g. \cite{Farkhooi2011}.
Unfortunately, the expressions obtained in this paper cannot be used for conclusions
made in terms of CC. This is because all expressions, including $p(t_1,t_0)$
are obtained under restriction (\ref{Domain}), whereas in order to calculate CC one needs to know
$p(t_1,t_0)$ for all $t_0>0$, $t_1>0$. Nevertheless, expressions derived in this paper 
under restriction (\ref{Domain})
allows one to show
that the Markov property is broken in the output ISI due to delayed feedback. Another reason
for neuronal activity to be non-Markov in a network is offered in \cite{Cessac2011}.

In further work, it is expected to extend obtained here exact expressions to the full range
of ISI values and to compare our findings with those obtained in terms of CC.
This includes also a quantitative estimation of how much the statistics is non-Markov
and to what extent it might be approximated by a Markow/renewal process. Also, a general
renewal stochastic process can be considered as a stimulus instead of Poisson one.
The latter can be achieved if to find adequate expression for Eq. (\ref{pisnit}), which
in its current form is valid for Poisson stimulation only.
\bigskip

{\small\bf Acknowledgements}
{\small This paper was partially supported by the Program
"Structure and Dynamics of Statistical and Quantum-Field Systems" of the National Academy of Science of
Ukraine, Project PK No 0117U000240.

Some calculations in the Appendix are made with the help of free 
Computer Algebra System Maxima, see \verb-http://maxima.sourceforge.net/-.
}

\appendix

\section{Appendix}

Here we give a simple example of the proven property. Namely, we consider
$p^{inh}(t_2\mid t_1,t_0)$ and show that $t_0$-dependence cannot be
eliminated for a LIF model, stimulated with Poisson stream.

For the two values of $n=1,2$, Eq. (\ref{pdfs}) due to (\ref{Domain}), (\ref{pts}), (\ref{intD})-(\ref{intDn+1}), (\ref{pn}), turns into the following two equations:
\begin{multline}\label{A1pn}
p^{inh}(t_1,t_0)=
p^0(t_1)\int\limits_0^{t_0}\mathbf{P}^0(s)p^0(t_0-s)g(s)ds+
\\
+p^0(t_0)\int\limits_0^{t_1}\mathbf{P}^0(s)p^0(t_1-s)g(s+t_0)ds+
\\
+p^0(t_1)p^0(t_0)\int\limits_0^{\Delta-t_0-t_1}f(s+t_0+t_1)ds,
\end{multline}
\begin{multline}\label{Apn+1}
p^{inh}(t_2,t_1,t_0)=
\\=
p^{inh}(t_2\mid \Delta-t_1)p^0(t_1)\int\limits_0^{t_0}\mathbf{P}^0(s)p^0(t_0-s)g(s)ds+
\\
+p^{inh}(t_2\mid \Delta)p^0(t_0)\int\limits_0^{t_1}\mathbf{P}^0(s)p^0(t_1-s)g(s+t_0)ds+
\\
+p^0(t_1)p^0(t_0)\int\limits_0^{\Delta-t_0-t_1}p^{inh}(t_2\mid s)f(s+t_0+t_1)ds,
\end{multline}

Now, let the neuronal model be the basic LIF model characterized with the firing
threshold $V_0$, input impulse height $h$ and relaxation time $\tau$. Assume that
\begin{equation}\label{Th=2}
0<h<V_0<2h.
\end{equation}
Assume also that the neuron is stimulated with a Poisson stream of intensity $\lambda$.
For this case, it is proven in (\cite[Eqs. (14),(21)]{Vidybida2014b} that
\begin{equation}\label{Ap0}
p^0(t)=\lambda^2te^{-\lambda t}, \text{ provided } t\in[0;T_2],
\end{equation}
where
$$
T_2=\tau\log\frac{h}{V_0-h}.
$$
Assume, for simplicity, that $\Delta<T_2$. This allows to obtain exact expressions
for $\mathbf{P}^0(s)$ and $p^{inh}(t\mid s)$:
\begin{equation}\label{AP0}
\mathbf{P}^0(s)=e^{-\lambda s}(\lambda s +1),
\end{equation}
$$
p^{inh}(t\mid s) = 
\lambda^2e^{-\lambda t}(t\chi(s-t) + \chi(t-s)(t-s)(\lambda s +1)).
$$
If we put $\chi(0)=0.5$, then $\chi(-x)=1-\chi(x)$ and the last expression can be transformed as follows
\begin{equation}\label{Apts}
p^{inh}(t\mid s) = 
\lambda^2e^{-\lambda t}(t+ \chi(t-s)s(\lambda(t-s) -1)).
\end{equation}

Under the assumptions of this Appendix, it appears that the kernel of integral equation
(\ref{trans}), above, is exactly the same as for the binding neuron model with excitatory
feedback. The latter case 
has been studied in  \cite[Eqs. (14)-(16)]{Vidybida2008a}, where the unique solution
to Eq. (\ref{trans}) is found. The unknown in general case quantities from Eq. (\ref{ff}),
above, under assumptions of this Appendix can be taken from \cite{Vidybida2008a}:
\begin{equation}\label{Ag}
g(s)=\frac{a\lambda}{2}\left(1-e^{-2\lambda(\Delta-s)}\right),
\end{equation}
\begin{equation}\label{Aa}
a=\frac{4e^{2\lambda\Delta}}{(2\lambda\Delta+3)e^{2\lambda\Delta}+1}.
\end{equation}
After substituting (\ref{Ap0})-(\ref{Aa}) into (\ref{A1pn}) one obtains
\begin{multline}\label{A1pnfinal}
p^{inh}(t_1,t_0)=\frac{\lambda^4 e^{-\lambda(t_0+t_1)}t_0t_1}{6((2\lambda\Delta+3)e^{2\lambda\Delta}+1)}\times
\\
\times\Big(2\lambda e^{2\lambda\Delta}
(\lambda(t_0^2+t_1^2) +3(2\Delta-t_0-t_1))+
\\
+3(e^{2\lambda(t_1+t_0)}+6e^{2\lambda\Delta}+1)\Big)\,.
\end{multline}
Notice, that $p^{inh}(t_1,t_0)$ is strictly positive for strictly positive $t_1$, $t_0$.
This allows one to use it safely as denominator in  the definition of conditional probability
(\ref{defcond}), above. Also, as it may be observed from (\ref{A1pnfinal}), the
 $p^{inh}(t_1,t_0)$ is continuous and bounded. This means, that $p^{inh}(t_2\mid t_1,t_0)$
 as it is defined in (\ref{defcond}), will preserve any discontinuity which may appear in
 the $p^{inh}(t_2,t_1,t_0)$, which is the numerator in (\ref{defcond}) for $n=1$.
 
\begin{figure}
\includegraphics[width=0.48\textwidth]{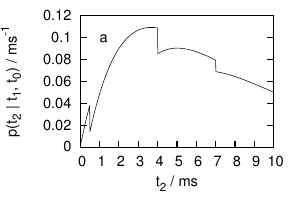}
\includegraphics[width=0.48\textwidth]{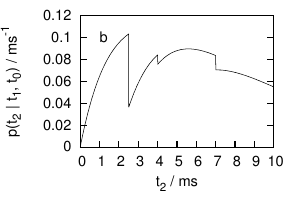}
\caption{\label{Fig}Different values of $p^{inh}(t_2\mid t_1, t_0)$ for different $t_0$.
Here $\Delta=7$ ms, 
$\lambda=0.3$ ms$^{-1}$, $t_1=3$ ms both for (a) and (b).
The $t_0=3.5$ ms for (a) and $t_0=1.5$ ms for (b).  
The curves are calculated based on Eqs. (\ref{Apts})-(\ref{Apn+1next}).
}
\end{figure}

 Consider now Eq.(\ref{Apn+1}) for $p^{inh}(t_2,t_1,t_0)$. After partial simplifications,
 it turns into the following:
\begin{multline*}            \tag{a}
 p^{inh}(t_2,t_1,t_0)=
 \\=
p^{inh}(t_2\mid\Delta-t_1)
\frac{\lambda^4 t_0t_1 e^{-\lambda (t_0+t_1)}}{6((2\lambda\Delta+3)e^{2\lambda\Delta}+1)}\times
\\\times
\left(3\left(1-e^{2\lambda t_0}\right) +2\lambda t_0 e^{2\lambda\Delta}(\lambda t_0+3)\right)+
\end{multline*}\vspace{-2\baselineskip}
\begin{multline*}            \tag{b}
+
p^{inh}(t_2\mid\Delta)
\frac{\lambda^4 t_0 t_1 e^{-\lambda(t_0+t_1)}}{6((2\lambda\Delta+3)e^{2\lambda\Delta}+1)}\times
\\\times
\left(3e^{2\lambda t_0}\left(1-e^{2\lambda t_1}\right) +2\lambda t_1 e^{2\lambda\Delta}(\lambda t_1+3)\right)+
\end{multline*}\vspace{-2\baselineskip}
\begin{multline*}            \tag{c}
+\lambda^2 t_2 e^{-\lambda t_2}p^0(t_1)p^0(t_0)\times
\\\times
\chi(\Delta-t_0-t_1-t_2)\int\limits_{t_2}^{\Delta-t_0-t_1}g(s+t_0+t_1)ds+
\end{multline*}\vspace{-2\baselineskip}
\begin{multline*}            \tag{d}
+
\lambda^2e^{-\lambda t_2}p^0(t_1)p^0(t_0)\times
\\\times
\int\limits_0^{min(t_2,\Delta-t_0-t_1)}(t_2-s)(\lambda s +1)g(s+t_0+t_1)ds+
\end{multline*}\vspace{-\baselineskip}
\begin{equation}\label{Apn+1next}
+a\,p^0(t_1)p^0(t_0) p^{inh}(t_2\mid \Delta-t_0-t_1).
\end{equation}
The summands (a) and (b) in Eq. (\ref{Apn+1next}) correspond to the first and second term of the right-hand side
in Eq.(\ref{Apn+1}), respectivly. The remaining three ones correspond to the the 
third term of the right-hand side in Eq.(\ref{Apn+1}). It is easily seen that both (a) and (b) are continuous
with respect to $t_0$. The term (d) is as well continuous in $t_0$, because function 
$\min(x,y)$ is continuous on $x$ and $y$.
The term (c) is as well continuous because
$$
\lim\limits_{\Delta-t_0-t_1-t_2\to 0}
\chi(\Delta-t_0-t_1-t_2)\!\!\int\limits_{t_2}^{\Delta-t_0-t_1}\!g(s+t_0+t_1)ds =0.
$$
Consider the final term in Eq. (\ref{Apn+1next}). For the sake of clarity, we omit
the factor $a\,p^0(t_1)p^0(t_0)$ having in mind that it is continuous and strictly positive
for $t_0>0$, $t_1>0$. The remaining expression is as follows
\begin{equation}\label{Aptsfinal}
\lambda^2e^{-\lambda t_2}
(t_2 - \chi(t_2+t_1+t_0-\Delta)(\Delta-t_0-t_1)(\lambda (\Delta-t_0-t_1-t_2)+1)).
\end{equation}
This expression, if considered as a function of $t_0,t_1,t_2$ has a step-like
discontinuity along the hyperplane
\begin{equation}\label{Ahyper}
t_2+t_1+t_0=\Delta.
\end{equation}
Indeed, if $t_2+t_1+t_0<\Delta$, then (\ref{Aptsfinal}) turns into
$$
\lambda^2e^{-\lambda t_2}t_2.
$$
Otherwise, if $t_2+t_1+t_0>\Delta$, then (\ref{Aptsfinal}) turns into
$$
\lambda^2e^{-\lambda t_2}
(t_2 - (\Delta-t_0-t_1)(\lambda (\Delta-t_0-t_1-t_2)+1)).
$$
The difference between the two expressions is as follows
\begin{equation}\label{Adiff}
\lambda^2e^{-\lambda t_2}
(\Delta-t_0-t_1)(\lambda (\Delta-t_0-t_1-t_2)+1).
\end{equation}
This difference vanishes along the hyperplane
$$
t_2+t_1+t_0=\Delta+\frac{1}{\lambda}
$$
only (due to (\ref{Domain}), we do not consider the case $t_0+t_1=\Delta$).
Comparing the last equation with (\ref{Ahyper}), we see that the jump (\ref{Adiff}) is
strictly positive along the hyperplane (\ref{Ahyper}).
The same is valid for the last term in (\ref{Apn+1next}).
Taking into account that the other four terms in (\ref{Apn+1next}) are continuous
in $t_0$, and what is said after Eq. (\ref{A1pnfinal}), we conclude
that $p^{inh}(t_2\mid t_1,t_0)$ has a nonzero jump along the hyperplane (\ref{Ahyper}).
For a fixed  $t_1,t_2$ and infinitesimally small $\epsilon>0$ consider two different
values of $t_0$: $t_0^{\pm}=\Delta-t_1-t_2\pm\epsilon$. It is clear from the above
that when $t_0$ value obtains infinitesimally small change from $t_0^+$ to $t_0^-$, the  $p^{inh}(t_2\mid t_1,t_0)$ gets finite change due to the jump (\ref{Adiff}), which means that
$t_0$-dependence in $p^{inh}(t_2\mid t_1,t_0)$ is indeed present 
(at least due to the discovered jump discontinuity)
 and cannot be eliminated.
This is illustrated in Fig. \ref{Fig}.




\bibliographystyle{ws-ijns}


\begin{thebibliography}{10}

\bibitem{Stein1967}
R.~B. Stein, Some models of neuronal variability, {\em Biophysical Journal}
  {\bf 7}(1)  (1967)  37--68.

\bibitem{Bryant1976}
H.~L. Bryant and J.~P. Segundo, Spike initiation by transmembrane current: a
  white-noise analysis, {\em The Journal of Physiology} {\bf 260}(2)  (1976)
  279--314.

\bibitem{Capocelli1971}
R.~M. Capocelli and L.~M. Ricciardi, Diffusion approximation and first passage
  time problem for a model neuron, {\em Kybernetik} {\bf 8}(6)  (1971)
  214--223.


\bibitem{Ghosh-Dastidar2009}
S.~Ghosh-Dastidar and H.~Adeli, Spiking neural networks, {\em International
  Journal of Neural Systems,} {\bf 19}(4)  (2009)  295--308.

\bibitem{Sacerdote2013}
L.~Sacerdote and M.~T. Giraudo, Stochastic integrate and fire models: A review
  on mathematical methods and their applications, {\em Stochastic
  Biomathematical Models, Lecture Notes in Mathematics 2058.\/},  eds.
  M.~Bachar, J.~J. Batzel and S.~Ditlevsen (Springer-Verlag, 2013), pp.
  99--148.

 \bibitem{Sarpeshkar2009}
R.~Sarpeshkar,
Neuromorphic and biomorphic engineering systems
{\em McGraw-Hill Yearbook of Science \& Technology 2009\/},
(New York: McGraw-Hill, 2009),
 pp. 250--252.
 
  \bibitem{Brderle2011}
 D.~Brüderle, M.A.~Petrovici, B.~Vogginger, B. et al.,
 A comprehensive workflow for general-purpose neural modeling with highly configurable neuromorphic hardware systems
{\em Biol Cybern,} {\bf 104}(4) (2011) 263--296.

\bibitem{Kravchuk2013}
K.~G. Kravchuk and A.~K. Vidybida, Firing statistics of inhibitory neuron with
  delayed feedback. II: Non-markovian behavior, {\em BioSystems} {\bf 112}(3)
  (2013)  233--248.


  
\bibitem{Vidybida2014}
A.~K. Vidybida, Binding neuron, {\em Encyclopedia of information science and
  technology\/},  ed. M.~Khosrow-Pour (IGI Global, 2014), pp. 1123--1134.


\bibitem{Lindner2010}
B. Lindner,
A brief introduction to some simple stochastic processes
{\em Stochastic Methods in Neuroscience\/}, 
ed. C.~Laing and G.J.~Lord (Oxford University Press, 2010), pp. 1--28.

\bibitem{Doob1953}
J.~L. Doob, {\em Stochastic processes} (Wiley, 1953).

\bibitem{vanKampen1998}
N.~G. van Kampen,
Remarks on Non-Markov Processes,
{\em Brazilian Journal of Physics} {\bf 28}(2) (1998) 90--96.


\bibitem{Vidybida2015}
A.~K. Vidybida, Activity of excitatory neuron with delayed feedback stimulated
  with Poisson stream is non-Markov, {\em Journal of Statistical Physics}
  {\bf 160} (2015)  1507--1518.


\bibitem{Burkitt}
A.~N. Burkitt, A review of the integrate-and-fire neuron model: I. homogeneous
  synaptic input, {\em Biological Cybernetics} {\bf 95}(1)  (2006)  1--19.

\bibitem{Chacron2003}
M.~J. Chacron, K.~Pakdaman and A.~Longtin, Interspike interval correlations,
  memory, adaptation, and refractoriness in a leaky integrate-and-fire model
  with threshold fatigue, {\em Neural Computation} {\bf 15}(2)  (2003)
  253--278.

\bibitem{Jolivet2004}
R.~Jolivet, T.~J. Lewis and W.~Gerstner, Generalized integrate-and-fire models
  of neuronal activity approximate spike trains of a detailed model to a high
  degree of accuracy, {\em Journal of Neurophysiology} {\bf 92}(2)  (2004)
  959--976.

\bibitem{Jolivet2006}
R.~Jolivet, A.~Rauch, H.~Lüscher and W.~Gerstner, Predicting spike timing of
  neocortical pyramidal neurons by simple threshold models, {\em J Comput
  Neurosci} {\bf 21}(1)  (2006)  35--49.

\bibitem{Vidybida2007}
O.~Vidybida, Output stream of a binding neuron, {\em Ukrainian Mathematical
  Journal} {\bf 59}(12)  (2007)  1819--1839.

\bibitem{Vidybida2014b}
O.~K. Vidybida, Output stream of leaky integrate and fire neuron, {\em Reports
  of the National Academy of Science of Ukraine} {\bf 2014}(12)  (2014)
  18--23.

\bibitem{Bekkers1998}
J.~M. Bekkers, Neurophysiology: Are autapses prodigal synapses?, {\em Current
  Biology} {\bf 8}(2)  (1998)  R52--R55.

\bibitem{Bacci2003}
A.~Bacci, J.~R. Huguenard and D.~A. Prince, Functional autaptic
  neurotransmission in fast-spiking interneurons: A novel form of feedback
  inhibition in the neocortex, {\em The Journal of Neuroscience} {\bf 23}(3)
  (2003)  859--866.

\bibitem{Bacci2004}
A.~Bacci, J.~R. Huguenard and D.~A. Prince, Long-lasting self-inhibition of
  neocortical interneurons mediated by endocannabinoids, {\em Nature} {\bf 431}
   (2004)  312--316.

\bibitem{Smith2002}
T.~C. Smith and C.~E. Jahr, Self-inhibition of olfactory bulb neurons, {\em
  Nature Neuroscience} {\bf 5}(8)  (2002)  760--766.

\bibitem{Benardo1994}
L.~S. Benardo, Separate activation of fast and slow inhibitory postsynaptic
  potentials in rat neocortex in vitro, {\em Journal of Physiology} {\bf 476.2}
   (1994)  203--215.

\bibitem{Storm1988a}
J.~F. Storm, Four voltage-dependent potassium currents in adult hippocampal
  pyramidal cells, {\em Biophys.J.} {\bf 53}  (1988) p. 148a.

\bibitem{Storm1990}
J.~Storm, Potassium currents in hippocampal pyramidal cells, {\em Progress in
  Brain Research} {\bf 83}  (1990)  161--187.

\bibitem{Rossell2012}
J.~L. Rosselló, V.~Canals, A.~Morro and A.~Oliver, Hardware implementation of
  stochastic spiking neural networks, {\em International Journal of Neural
  Systems} {\bf 22}(4)  (2012) p. 1250014.

\bibitem{Wang2013}
R.~Wang, G.~Cohen, K.~M. Stiefel, T.~J. Hamilton, J.~Tapson and A.~van Schaik,
  An FPGA implementation of a polychronous spiking neural network with delay
  adaptation, {\em Frontiers in Neuroscience} {\bf 7}(14)  (2013).

\bibitem{Vidybida2013}
A.~K. Vidybida and K.~G. Kravchuk, Firing statistics of inhibitory neuron with
  delayed feedback. I. output ISI probability density, {\em BioSystems} {\bf
  112}(3)  (2013)  224--232.

\bibitem{Vidybida2008a}
A.~K. Vidybida, Output stream of binding neuron with delayed feedback, {\em
  14th International Congress of Cybernetics and Systems of WOSC, Wroclaw,
  Poland, September 9-12, 2008\/},  eds. J.~Józefczyk, W.~Thomas and
  M.~Turowska (Oficyna Wydawnicza Politechniki Wroclawskiej, 2008), pp.
  292--302.

\bibitem{Averbeck2009}
B.~B. Averbeck, Poisson or not Poisson: Differences in spike train statistics
  between parietal cortical areas, {\em Neuron} {\bf 62}(3)  (2009)  310--311.

\bibitem{Lowen1992}
S.~B. Lowen and M.~C. Teich, Auditory-nerve action potentials form a nonrenewal
  point process over short as well as long time scales, {\em The Journal of the
  Acoustical Society of America} {\bf 92}(2)  (1992)  803--806.

\bibitem{Shinomoto1999}
S.~Shinomoto, Y.~Sakai and S.~Funahashi, The Ornstein-Uhlenbeck process does
  not reproduce spiking statistics of cortical neurons, {\em Neural
  Computation} {\bf 11}  (1999)  935--951.

\bibitem{Ratnam2000}
R.~Ratnam and M.~E. Nelson, Nonrenewal statistics of electrosensory afferent
  spike trains: implications for the detection of weak sensory signals, {\em
  The Journal of Neuroscience} {\bf 20}(17)  (2000)  6672--6683.

\bibitem{Nawrot2007}
M.~P. Nawrot, C.~Boucsein, V.~Rodriguez-Molina, A.~Aertsen, S.~Grün and
  S.~Rotter, Serial interval statistics of spontaneous activity in cortical
  neurons in vivo and in vitro, {\em Neurocomputing} {\bf 70}  (2007)
  1717--1722.

\bibitem{Maimon2009}
G.~Maimon and J.~A. Assad, Beyond Poisson: Increased spike-time regularity
  across primate parietal cortex, {\em Neuron} {\bf 62}(3)  (2009)  426--440.

\bibitem{Rospars1993a}
J.~P. Rospars and P.~L\'ansk\'y, Stochastic model neuron without resetting of
  dendritic potential: application to the olfactory system, {\em Biol. Cybern.}
  {\bf 69}(4)  (1993)  283--294.


\bibitem{Lnsk1999}
P.~L\'ansk\'y and R.~Rodriguez, Two-compartment stochastic model of a neuron, {\em
  Physica D: Nonlinear Phenomena} {\bf 132}(1–2)  (1999)  267--286.

\bibitem{Benedetto2013}
E.~Benedetto and L.~Sacerdote, On dependency properties of the ISIs generated
  by a two-compartmental neuronal model, {\em Biol Cybern} {\bf 107}(1)  (2013)
   95--106.


\bibitem{Kass2005}
R.~E. Kass, V.~Ventura and E.~N. Brown, Statistical issues in the analysis of
  neuronal data, {\em Journal of Neurophysiology} {\bf 94}(1)  (2005)  8--25.
  
\bibitem{Avila-Akerberg2011}
O.~Avila-Akerberg and M.~J. Chacron, Nonrenewal spike train statistics: causes
  and functional consequences on neural coding, {\em Experimental Brain
  Research} {\bf 210}(3-4)  (2011)  353--371.

\bibitem{Farkhooi2011}
F.~Farkhooi, E.~Muller and M.P.~Nawrot, Adaptation reduces variability of the neuronal population code,
{\em Phys Rev} {\bf E 83}(5) (2011) : 050905.

\bibitem{Cessac2011}
B.~Cessac, A discrete time neural network model with spiking neurons: II:
  Dynamics with noise, {\em Journal of Mathematical Biology} {\bf 62}(6)
  (2011)  863--900.

\end{thebibliography}


\end{document}